\documentclass[12pt]{article}
\usepackage{epsfig}
\def\eq#1{{(\ref{#1})}}
\def\fig#1{{Fig.~\ref{#1}}}
\def\re#1{{Ref.~\cite{#1}}}
\def\order#1{\mathcal{O}{(#1)}}
\newcommand{\beq}{\begin{equation}}
\newcommand{\eeq}{\end{equation}}
\newcommand{\beqar}[1]{\begin{eqnarray}\label{#1}}
\newcommand{\eeqar}{\end{eqnarray}}

\newcommand{\as}{\alpha_s}
\newcommand{\un}{\underline}

\newcommand{\stackeven}[2]{{{}_{\displaystyle{#1}}\atop\displaystyle{#2}}}
\newcommand{\lsim}{\stackeven{<}{\sim}}

\newcommand{\kg}{k_\mathrm{geom}}
\begin{document}

\title {
\begin{flushright}
{\small BNL--NT--03/34}\\[-0.3cm]
{\small INT--PUB--03--19}\\[-0.3cm]
{\small October, 2003}\\[0.5cm]
\end{flushright}
{Open Charm Production in Heavy Ion Collisions \\ and the Color Glass Condensate }}
\author{
{ Dmitri Kharzeev$\, ^a$  \quad and \quad  Kirill Tuchin$\, ^{a,b}$
} \\[5mm]
{\it\small $^a\,$Physics Department, Brookhaven National Laboratory,}\\
{\it\small Upton, NY 11973-5000, USA}\\[0.5cm]
{\it\small $^{b}\,$Institute for Nuclear Theory, University of 
Washington, 
Box 351550}\\
{\it\small Seattle, WA 98195, USA}
}

\date{}

\maketitle

\begin{abstract} 
We consider the production of open charm in heavy ion collisions in the
framework of the Color Glass Condensate. In the central rapidity region at RHIC, 
for the charm quark yield we expect $N_\mathrm{coll}$ (number of collisions) 
scaling in the 
absence of final--state effects.
At higher energies, or forward rapidities at RHIC, the saturation scale exceeds the charm 
quark mass; we find that this results in the approximate $N_\mathrm{part}$ (number of participants) scaling of charm production 
in $AA$ collisions and  $\sqrt{N_\mathrm{part}^A}$ scaling in $p(d)A$ collisions, similarly to the 
production of high $p_T$ gluons discussed earlier.
  We also show that 
the saturation phenomenon makes spectra harder as compared to the
naive parton model approach.  
We then discuss the energy loss of charm quarks in hot and cold media and 
argue that the hardness of the spectrum implies very slow dependence of the
quenching factor on $p_T$.

\end{abstract}

%%%%%%%%%%%%%%%%%%%%%%%%%%%%%%%%%%%%%%%%%%%%%%%%%%%%%%%%%%%

\section{Introduction}

Heavy quarks play a very important role in the development of Quantum
Chromodynamics.  The masses of heavy quarks $m$ significantly exceed the
QCD scale $\Lambda_\mathrm{QCD}$, which makes perturbative calculations of
charm production and annihilation possible \cite{collcharm}.  The QCD
decoupling theorems \cite{Appelquist:tg} ensure that heavy quarks do not
influence the dynamics of processes at scales much smaller than $m$.
  
At sufficiently small Bjorken $x$ and/or for sufficiently heavy nucleus
$A$, parton distributions approach saturation \cite{GLR}. In the saturation regime
partons form the Color Glass Condensate (CGC) characterized by the
dimensionful ``saturation scale'' $Q_s$ determined by the parton density 
\cite{GLR,Mueller:wy,Blaizot:nc,MV}.  
Possible manifestations of the Color Glass Condensate in the energy, centrality, 
and rapidity dependence of hadron multiplicities  \cite{KN,KL,KLM}
have been found by experiments at RHIC 
\cite{brahms,phenix,phobos,star}. 

It is intuitively clear that once the parton density becomes high enough
to ensure that $Q_s$ and $m$ are of the same order, heavy quarks will no
longer decouple. This leads to very interesting consequences for their
production dynamics, as we will discuss at some length in this paper.
First, let us however illustrate this point by using very simple
qualitative arguments.

The Color Glass Condensate is characterized by strong classical color 
fields \cite{MV,YuK}; 
the strength of the 
chromo--electric field can be estimated as
\beq \label{Efield}
E \sim {Q_s^2 \over g},
\eeq
where $g^2 = 4 \pi \alpha_s$ is the strong coupling constant. This field can polarize  
the Dirac vacuum of quarks; in particular, quark pairs can be produced when 
the potential difference across the Compton wavelength 
of the heavy quark becomes equal to the energy needed for the pair production,
\beq
gE \sim {m \over 1/m} = m^2.
\eeq
Together with (\ref{Efield}) this condition suggests that when 
\beq
Q_s^2 \geq m^2,
\eeq
the heavy quarks will no longer decouple, and their production pattern 
will be similar to that of light quarks.

It has been argued before \cite{KLM} that Color Glass Condensate at sufficiently high energy and/or rapidity 
leads to the suppression of high $p_T$ parton production. Namely, instead of the $N_{coll}$ (number of collisions) 
scaling of yields of 
high $p_T$ partons  
expected on the basis 
of QCD collinear factorization, one finds $N_{part}$ (number of participants) scaling in $AA$ collisions and   $\sqrt{N_\mathrm{part}^A}$ 
scaling in $p(d)A$ collisions. In view of the argument given above, one expects that the yields of heavy quarks can follow 
similar scaling. The goal of the present paper is to establish in what kinematical region this scaling can hold for heavy quark production. 

The present d-Au data \cite{dAdata} at mid-rapidity at RHIC do not support the 
suggestion \cite{KLM} that 
quantum evolution in the CGC is responsible for the observed in $Au-Au$ collisions suppression of high $p_T$ hadrons. 
Instead, they indicate the dominance of final state effects consistent with jet 
quenching \cite{GW,BDMPS,WS}. 
Nevertheless, at sufficiently large energy and/or rapidity the arguments of \cite{KLM} should apply. 
Since the saturation scale exponentially increases with rapidity, one may expect that this can happen already 
at RHIC energies in the forward rapidity region.
 
\vskip0.3cm
%\vskip0.3cm

Production of particles in relativistic heavy-ion collision is a 
complicated process which in the CGC framework can be considered as consisting of two stages.  
First, upon the collision partons are released from the wave functions of the nuclei.
At the quasi-classical level the wave function of the ultra-relativistic
nucleus is described by the non-abelian Weizs\"acker-Williams field
\cite{MV,YuK} created by the valence quarks moving along the light cone. 
Quantum
evolution of the wave function of the nucleus is described in the framework of the
Color Glass Condensate by a set of coupled evolution equations\cite{jmwlk}
which can be reduced to a single nonlinear equation in large $N_c$
approximation \cite{BK}. At the second stage, there are final state interactions 
between the produced partons which may bring the quark-gluon system to 
thermal equilibrium \cite{MSon}. In this paper we
address both problems in the case of the open charm production. Very recently, heavy quark production 
in the CGC framework has been considered also in Ref.\cite{GV}. Our treatment of the problem includes the effects of 
quantum evolution in the nuclear wave functions; they 
appear to be significant and lead to non-trivial centrality dependence of the heavy quark 
yields.

\section{Heavy quark production in $k_T$-factorization}

\subsection{Collinear and $k_T$ factorization schemes}

Unlike the inclusive gluon production case \cite{GLR,GyMc,KL,KLM}, as we discussed above the
heavy quark production is characterized by two inherent scales: the
saturation scale $Q_s$ and the heavy
quark mass $m$ (we assume that the two colliding nuclei have the same 
atomic number $A$).
Depending on the relation between these two scales we have two different
mechanisms of the quark production. The quasi-classical gluon field of a
nucleus is characterized be the unintegrated gluon distribution
$\varphi_A(k_\bot,y)$ which is flat at $k_\bot^2\le Q_s^2$\footnote{See
extensive discussion in \cite{KKT}.} and decreases as $1/k_\bot^2$
otherwise. A quark production amplitude $\mathcal{A}$ in the process
$gg\rightarrow q\bar q X$ has typical momentum $p_\bot^2\sim
\max\{m^2,Q_s^2\}$. The cross section of the quark production is given by
the convolution of the unintegrated gluon distributions of nuclei with the
quark production amplitude. This statement is known as $k_T$ factorization
\cite{LRSS,CCH,CE}. It was proved only in the case when multiple 
rescatterings
of partons from the nuclei wave functions are suppressed relative to the
evolution effects. Although the $k_T$ factorization has not been proven
for general process in high energy QCD, its numerous applications are
proved to be rather successful (for charm production, see e.g. \cite{apple}). 
In the context of relativistic heavy ion
program it was utilized in \cite{KL,KLM} to calculate the inclusive gluon
production. Additional confidence in $k_T$-factorization arises from the
fact that it holds exactly (includes all rescatterings in a nucleus) for
the gluon production in $\gamma^*A$ \cite{KTin}. This result also holds
for a $pA$ process at not very high energies, when a proton can be
considered as a diluted object \cite{KKT}.  In the framework of the CGC
$k_T$ factorization for the heavy quark production was recently demonstrated 
 in
\cite{GV}. In the following we will assume that the $k_T$-factorization
gives fairly good approximation at RHIC energies in the central rapidity
region even if the rescatterings are turned in.

The basic assumption of the collinear factorization is that the typical 
transverse momentum associated with  the colliding hadron is of the 
order of the $\Lambda_\mathrm{QCD}$ and much smaller than the typical 
momentum of the hard subprocess. In the case of heavy flavor production, 
large quark mass $m$ seems to insure the validity of the collinear 
factorization. However, at high energy the typical scale of the hadron 
wave function is $Q_s(y)$ which 
increases with energy. It becomes no longer possible to integrate out the 
transverse degrees of freedom of the hadron wave function. The 
$k_T$-factorization is thus the generalization of the collinear 
factorization at high energies.
When $Q_s\ll m$ a quark production amplitude
overlaps only with perturbative ($p_T\gg Q_s$) low density tail of
$\varphi$. Perturbative approach must be valid in that 
kinematic region although we will comment later how strong must the above
inequality be. In the opposite case $Q_s\gg m$
unintegrated
gluon distributions and the quark production amplitude strongly
overlap. The hard transverse momentum of gluons in a nucleus field can
be transmitted to the quark--anti-quark system giving rise to a strong
deviation from the naive perturbative approach. In particular, the
high parton density effects might manifest themselves in the 
form of the charmed-meson spectrum, dependence of
the quark multiplicity on the atomic weight $A$, angular asymmetry 
of quark jets production and overall enhancement of the quark
production cross section.

%%%%%%%%%%%%%%%%%%%%%%%%%%%%%%%
\subsection{$N_\mathrm{part}$ versus $N_\mathrm{coll}$ scaling in 
inclusive gluon production}
The question about the relation between scales $Q_s$ and $m$ is
closely related to the question about the dependence of the quark
multiplicity on the atomic weight $A$. Let us first review the
argument given for the inclusive hadron production case \cite{KL,KLM,KKT}. The
multiplicity spectrum is given by 
\beq\label{incl}
\frac{dN}{dy d^2p_\bot}=\frac{2\pi\as}{C_F\,S_A}\,\frac{1}{p_\bot^2}
\int dk_\bot^2 \varphi_A(x_1,k_\bot^2)\,\varphi_A(x_2, (p-k)_\bot^2).
\eeq
We would like to consider it in three kinematic regions which are
defined by three dimensional parameters $Q_0$, $Q_s$ and  
$\kg= Q_s^2/Q_0$. Here $Q_0$ is the characteristic scale of 
nuclear
hadronization region. The scale $\kg$ sets the region of
applicability of the collinear factorization. Indeed, at $k_\bot\ll
\kg$ we have (modulo logarithms)
\beq\label{coll}
\ln k_\bot^2/Q_0^2\,=\, \ln k_\bot^2/Q_s^2\,+\,  \ln Q_s^2/Q_0^2
\,\approx\,\ln Q_s^2/Q_0^2\,=\, \lambda\,y,
\eeq
where $\lambda\sim \mathcal{O}(\alpha_s)$. 
Thus, the evolution with respect to the longitudinal momentum (BFKL) cannot
be neglected with respect to the evolution in the transverse momentum
(DGLAP) despite the fact that evolution is linear. We will estimate
the integral in \eq{incl} assuming that $k_\bot^2\ll p_\bot^2$. This
corresponds to the logarithmic (in $p_\bot$) approximation. Note, that
integration in \eq{incl} goes always over the smallest transverse momentum
$\min\{k_\bot^2,(p-k)_\bot^2\}$, while the external one is the
largest. This can be easily seen by rewriting \eq{incl} in symmetric
from  in terms of $k_\bot$ and $q_\bot=p_\bot-k_\bot$.
In the region of applicability of the parton model $p_\bot^2>
\kg^2$ we have
\beq\label{reg1}
\frac{dN}{dy d^2p_\bot}\,\sim\, 
\frac{S_A}{\as\,p_\bot^2}\,\int^{p_\bot^2}
dk_\bot^2\,\frac{Q_s^2}{k_\bot^2}\,\frac{Q_s^2}{p_\bot^2}\, 
\sim\, S_A Q_s^4\,\frac{1}{\as\,p_\bot^4}\sim N_\mathrm{coll},
\eeq
since $Q_s^2\sim A^{1/3}$ and $S_A\sim A^{2/3}$. 
In the region   $Q_s^2\le p_\bot^2\ll \kg^2$ the evolution is
linear but existence of the strong color field at long
distances affects the gluon distribution which acquires the anomalous
dimension $1/2$: $\varphi\sim k_\bot/Q_s$ \cite{LT,IIM,MT}. Thus,
\beq\label{reg2}
\frac{dN}{dy d^2p_\bot}\,\sim\, 
\frac{S_A}{\as\,p_\bot^2}\,\int^{p_\bot^2}
dk_\bot^2\,\left(\frac{Q_s^2}{k_\bot^2}\right)^{1/2}\,
\left(\frac{Q_s^2}{p_\bot^2}\right)^{1/2}\, 
\sim\, S_A Q_s^2\,\frac{1}{\as\,p_\bot^2}\sim N_\mathrm{part}
\eeq
which is known as the geometric scaling \cite{geom}.
Finally, in the saturation region   $ p_\bot^2\ll Q^2_s$
\beq\label{reg3}
\frac{dN}{dy d^2p_\bot}\,\sim\, 
\frac{S_A}{\as\,p_\bot^2}\,\int^{Q_s^2}
dk_\bot^2\, 
\,\sim\, S_A Q_s^2\,\frac{1}{\as\,p_\bot^2}\sim N_\mathrm{part}
\eeq
We reproduce the result of \cite{KLM} that the multiplicity scale with $
N_\mathrm{part}$ in a vast kinematic region $k_\bot < \kg$.  Similar considerations lead 
to $\sqrt{N_{part}^A}$ scaling in $p(d)A$ collisions for sufficiently central collisions.

The transition from the classical regime dominated by the Cronin effect \cite{JNV} to the suppression regime 
was discussed in our recent paper with Yu. Kovchegov 
\cite{KKT}. 
This issue was addressed recently in Refs.~\cite{KW}, which confirmed the phenomenon 
of high $p_T$ suppression at sufficiently large rapidity and studied its evolution.

%%%%%%%%%%%%%%%%%%%%%%%%%%%%%%%%%
\subsection{Heavy quark production}
Returning to the case of heavy quark production,  the 
saturation corresponds to the region $m_\bot < Q_s$, where 
$m_\bot^2=p_\bot^2+m^2$. To estimate the scale of the collinear
factorization breakdown we note that the heavy quark threshold in the
$s$-channel is $4m^2$. Analogously to the gluon production case 
considered above, the scale at which collinear
factorization breaks down is 
%\beq\label{qtc}
%\kg^{c\,2}\,=\,\frac{Q_s^4}{Q_0^2}\,-\,4\,m^2.
%\eeq 
\beq\label{qtc}
\mathcal{M} \,=\,\frac{Q_s^2}{Q_0},
\eeq
where $\mathcal{M}$ is the invariant mass of the heavy quark pair; for 
the quarks 
produced at close rapidities, $\mathcal{M}^2 \simeq 4\, (p_{\bot}^2 + 
m^2)$ (in their CM system).
Since at  RHIC $Q_s$ is of the same order as $m$ in the midrapidity, 
\eq{qtc} is not satisfied even at $p_\bot =0$. Therefore, 
 we expect to
observe the $N_\mathrm{coll}$ scaling of the open charm spectrum. 
At the forward rapidity $Q_s\gg m$ and we expect to observe 
$N_\mathrm{part}$ scaling for momenta $p_\bot^2 < \max\{Q_s^2, \,
Q_s^4/(4Q_0^2)-m^2\}$. In any case, the 
charmed-meson spectrum is expected to be much harder than that 
predicted by the parton model, since the typical gluon momentum in the 
nucleus wave function is $Q_s\gg \Lambda_\mathrm{QCD}$. The total cross 
section is also expected to be higher since there are additional 
contributions to the production amplitude as compared to the parton model.

It was argued in Refs.~\cite{LRSS,CCH,CE} that the cross section of the
hadroproduction of the heavy quarks at high energies can be written in
a $k_T$-factorized form:
\begin{eqnarray}
\frac{d\sigma}{d^2 p_{\bot 1}\, dy_1^*dy_2^*}&=&
\int \frac{d^2 q_{\bot 1}}{\pi} \int \frac{d^2 p_{\bot 2}}{\pi}\,
\varphi_A (y, q_{\bot 1}^2)\, 
\varphi_A(x,  (p_{\bot 1}+ p_{\bot 2}- q_{\bot 1})^2)\,\nonumber\\
&&\times
\frac{\alpha_s^2\,}{8 \pi^2 x^2 y^2 S^2}\,
\mathcal{A}(gg\rightarrow ggc\bar c),
\label{pp}
\end{eqnarray}
Here $\un p_1$, $\un p_2$ and $y_1^*$, $y_2^*$ 
are the quark and antiquark transverse momenta and rapidities,
$S=(p_A+p_B)^2$ and 
\begin{eqnarray}
x_{1,2} &=& (m_{\bot 1,2}^2/S)^{1/2}\,e^{-y_{1,2}^*},\quad
x\,=\,x_1\,+\,x_2\label{xy1}\\
y_{1,2} &=& (m_{\bot 1,2}^2/S)^{1/2}\,e^{y_{1,2}^*},\quad
y\,=\,y_1\,+\,y_2,\label{xy2}
\end{eqnarray}
$\varphi_A(x, q^2_\bot)$ is unintegrated gluon distribution in a proton
defined as
\beq\label{distr}
\varphi_A(x, q^2_\bot)\,=\, \frac{d\, xG_A(x, q^2_\bot)}{d q^2_\bot}.
\eeq
$\mathcal{A}(gg\rightarrow ggc\bar c)$ 
is a production amplitude the exact expression for
which is rather nasty and can be found in the Appendix.
In the limit of vanishing virtualities of the $t$-channel
gluons $\un q_1^2$, $\un q_2^2 \rightarrow 0$ equation \eq{pp} reduces
to the well known parton model expression (collinear factorization)
\cite{CCH}.
%%%% fig %%%%%%%%
\begin{figure}
\begin{center}
\epsfig{file=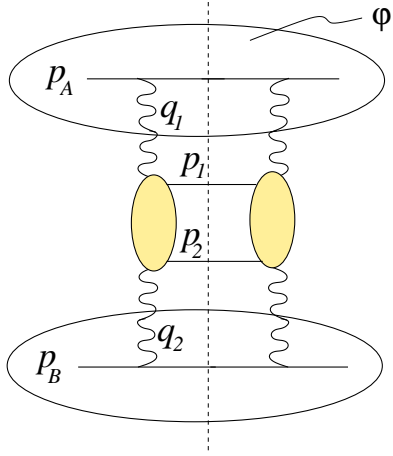, height=6cm}
\caption{Production of $q\bar q$ pair in heavy-ion collisions in
  $k_T$-factorization approach. 
}\label{ppcc}
\end{center}
\end{figure}
%%%%%%%%%%%%

%%%%%%%%%%%%%%%%%%%%%%%%%%%%%%%%%
%\subsection{$N_\mathrm{part}$ versus $N_\mathrm{coll}$ scaling in 
%heavy quark production}
The amplitude \eq{amp} is rather hard for analytical analysis. We can 
simplify it in the Regge kinematics $|y_1^*-y_2^*|\gg 1$.
This limit cannot be used in numerical calculations since
quarks tend to be produced at the same rapidity: production of large 
invariant masses of the $q\bar q$ pair is suppressed. 
However it gives a fair qualitative understanding of the underlying 
physics.
A simple calculation yields
\beq\label{multi}
\frac{1}{x^2y^2S^2}\,\mathcal{A}(gg\rightarrow ggc\bar c)\approx
\frac{e^{-\Delta y^*}}{2m_{1\bot}m_{2\bot}N_c}\,
\frac{1}{(q_1-p_1)_\bot^2+m^2}\,+\,\order{e^{-2\Delta y^*}}.
\eeq
In the region of applicability of collinear factorization, 
$m_{1\bot}\rightarrow \infty $,  which 
at RHIC is the same as $m_{1\bot}\gg Q_{s\mathrm{max}}$, 
the transverse momenta of gluons are negligible compared to
those of produced heavy quarks. Therefore $p_{1\bot}\approx -p_{2\bot}$ 
and \eq{pp}
can be written as (we omit numerical factors)
\begin{eqnarray}
\frac{dN^\mathrm{c\bar c}}{d^2 p_{\bot 1}\, dy_1^*dy_2^*}&\propto&
\frac{\as^2 e^{-\Delta y^*}}{S_A}
\int^{p_{1\bot}^2} \frac{dp_{2\bot}^2}{m_{1\bot}m_{2\bot}}
\int\, dq_{1\bot}^2
\frac{1}{N_c\,p_{1\bot}^2}
\frac{A^2\,xG_N(q_{1\bot}^2)\,
xG_N(p_{2\bot}^2)}{q_{1\bot}^2\,p_{2\bot}^2}
\nonumber\\
&\sim&
\frac{\as^2\,A^2\, e^{-\Delta y^*} 
(xG_N(p_{1\bot}^2))^2}{N_c\, S_A\, p_{1\bot}^4}\,\sim\,
 N_\mathrm{coll},\label{obl1}
\end{eqnarray}
where clearly $p_{1\bot}^2\gg m^2$. Note that the
exponential suppression at large $\Delta y^*$ of the heavy quark
amplitude as compared to the $gg\rightarrow gggg$ \cite{LO} is because
$t$-channel quark carries spin $1/2$. 

At large rapidities there is a kinematic region such that 
$Q_{s\mathrm{min}}
\ll m_{1\bot}\ll Q_{s\mathrm{max}}$. In this case one of the nuclear
wave functions is in the saturation whereas another one is not. In
general transverse momenta of quarks are no longer equal and point
back-to-back. Thus in the limit $p_{1\bot}^2\gg p_{2\bot}^2$ the quark
multiplicity reads
\begin{eqnarray}
\frac{dN^\mathrm{c\bar c}}{d^2 p_{\bot 1}\, dy_1^*dy_2^*}&\propto&
\frac{\as^2 e^{-\Delta y^*}}{S_A}
\int \frac{dp_{2\bot}^2}{m_{1\bot}m_{2\bot}}
\int^{Q_{s\mathrm{max}}^2}\, dq_{1\bot}^2
\frac{1}{N_c\,p_{1\bot}^2}\frac{S_A\,Q_{s\mathrm{min}}^2}{q_{1\bot}^2}\,
\frac{S_A}{\as}\nonumber\\
&\sim& \frac{1}{N_c}
\frac{ e^{-\Delta y^*}\, S_A\,Q_{s\mathrm{min}}^2}
{p_{1\bot}^2}\,\ln\frac{Q_{s\mathrm{max}}^2}{Q_{s\mathrm{min}}^2}
\,\sim\, N_\mathrm{part}\, \ln\frac{Q_{s\mathrm{max}}^2}{Q_{s\mathrm{min}}^2}.
\label{obl2}
\end{eqnarray}
Please, note that here $Q^2_{s\mathrm{min}}\sim \as A/S_A$.

Finally, there is a region where both nuclei are in saturation 
$m_{1\bot}\ll Q_{s\mathrm{min}}$. We have
\begin{eqnarray}
\frac{dN^\mathrm{c\bar c}}{d^2 p_{\bot 1}\, dy_1^*dy_2^*}&\propto&
\frac{\as^2 e^{-\Delta y^*}}{S_A}
\int \frac{dp_{2\bot}^2}{m_{1\bot}m_{2\bot}}
\int^{Q_{s\mathrm{min}}^2}\, dq_{1\bot}^2
\frac{1}{N_c\,p_{1\bot}^2}\frac{S_A^2}{\as^2}\nonumber\\
&\sim& \frac{1}{N_c}
\frac{e^{-\Delta y^*}\, S_A\,Q_{s\mathrm{min}}^2}
{p_{1\bot}^2}
\,\sim\, N_\mathrm{part}.\label{obl3}
\end{eqnarray}

Consider now the total multiplicity of heavy quark pairs produced in
heavy ion collisions per unit rapidity, $dN^\mathrm{c\bar c}/ dy_1^*dy_2^*$.
At RHIC in the kinematic region $|y^*| \lsim 1$ we have 
$Q_s^2(y^*=0)\simeq 1-2$ GeV$^2$. Since $m\simeq 1.2-1.9$~GeV \cite{PDG} 
the
transverse size of charmed quark $1/m$ is too small to feel saturation
in either one of the nuclei and set $N_\mathrm{part}$ scaling. The 
parametric dependence of 
the total multiplicity of heavy quark pairs is (see \eq{obl1}) 
\beq\label{thum1}
\frac{dN^\mathrm{c\bar c}}{ dy_1^*dy_2^*}\,\sim\,
\frac{\as^2 A^2}{S_A}\sim N_\mathrm{coll},\quad y^*\simeq 0
,\quad \mathrm{RHIC}.
\eeq
Nevertheless, the  $k_T$-factorization result for total multiplicity is 
numerically different from the parton model one: since in 
$k_T$-factorization spectrum is harder we get larger total cross section.
 
Note, that since at RHIC  $Q_s\sim m$ the variation of the
saturation scale with rapidity becomes essential. 
Variation of $Q_s$ with rapidity stems
from the fact that the saturation scale is proportional to the gluon
density in the transverse plane of nucleus, which  in turn is
proportional to the gluon structure function $xG(x,k_\bot^2)$. 
Since $xG(x,k_\bot^2)$ is increasing  function of rapidity
$y^*=\ln(1/x)$ we conclude that $Q_s$ grows with $y^*$ defined with
respect to the rapidity of the nucleus in a given reference frame.
Therefore, relation between saturation scales of each nucleus and
the quark mass are different at different rapidities $y^*$.  
If we consider quarks produced at $|y^*|\gg 1$ (but still not in
hadronization region), then one of the nuclei is in saturation while
another one is not. In that case we estimate using \eq{obl2}
\beq\label{thum2}
\frac{dN^\mathrm{c\bar c}}{ dy_1^*dy_2^*}\,\sim\,
\as A\sim N_\mathrm{part},\quad |y^*|\gg 1,\quad \mathrm{RHIC,\; LHC}.
\eeq
In $p(d)A$ collisions, we expect that in the deuteron fragmentation region and for sufficiently 
central events, the scaling will be 
\beq\label{thum2a}
\frac{dN^\mathrm{c\bar c}}{ dy_1^*dy_2^*}\,
\sim \sqrt{N_\mathrm{part}^A},\quad |y^*|\gg 1,\quad \mathrm{RHIC,\; LHC}.
\eeq
In the next section we will quantify the kinematical region at which the transition to 
the new scaling will take place at RHIC.

At LHC energies the situation is different. We expect that the
saturation scale in the midrapidity region will be much larger than
the charmed quark mass. As the result both nuclei will be in the
saturation. By \eq{obl3} we get
\beq\label{thum3}
\frac{dN^\mathrm{c\bar c}}{dy_1^*dy_2^*}\,\sim\,
\as^0 A\sim N_\mathrm{part},\quad y^*\simeq 0,\quad \mathrm{LHC},
\eeq
whereas far from midrapidity one of the nuclei is not in saturation and
we can use \eq{thum2}. Note that from \eq{xy1},\eq{xy2} and \eq{QSC}
it follows that the ratio between the largest and the smallest
saturation scales is
\beq\label{ratio}
\frac{Q^2_{s\mathrm{max}}}{Q^2_{s\mathrm{min}}}\,\simeq\,
e^{\lambda(|y_1^*|+|y_2^*|)}.
\eeq
This ratio is much larger than unity when at least
one of the quarks is produced at $|y^*|\gg 1$ although in most cases
quarks are produced at $|y_1^*-y_2^*|\lsim 1$ since otherwise the
amplitude is exponentially suppressed (see \eq{multi}).

%%%%%%%%%%%%%%%%%%%%%%%%%%%%%%%%%%%%%%%%%%%%%%%%%%%%%%%%%%%%%%%%%
\section{Open charm production in $AA$ and $dA$ collisions} 

\subsection{A model for unintegrated gluon distribution} 
To proceed
further we have to specify the unintegrated gluon distribution of a
nucleus $\varphi_A(x,\un q)$. In principle it is directly related to the
forward elastic scattering amplitude, which in turn can be calculated from 
the
nonlinear evolution equations \cite{GLR,BK}. However, their exact 
analytical solution is
not known. Although recently the progress in obtaining the numerical
solution was reported by many authors \cite{LL} we prefer to use simple
parametrization which catches the most essential details of the solution.
In this paper we employ a model for distribution function suggested in
\cite{KLM}. It matches the known analytical expressions in the asymptotic
regions \cite{LT,Iancu:2001md}. It reads
\begin{eqnarray}
\varphi_A(x, q^2_\bot) &=&
\frac{4\,C_F}{2(2\pi)^2}\frac{S_\bot}{\alpha_s}\, d\left(\frac{
  q^2_\bot}{Q_s^2(y^*)}\right)\,
(1-x)^4\,\Bigg[\theta(Q_s^2\,-\,q_\bot^2)
\nonumber\\ 
&& +\,  \left(\frac{Q_s^2(y^*)}{ q^2_\bot}\right)^{1.3\,
  \alpha_s}\,
I_0\left\{\left(\ln\frac{Q_s^2(y^*)}{Q_s^2(y_0^*)}\,\ln\frac{
q^2_\bot}{Q_s^2(y_0^*)}\right)^{1/2}\right\}\,\theta(q^2_\bot\,-\,Q_s^2)
\Bigg]
\label{dist}
\end{eqnarray}
The modified Bessel function $I_0$ is a solution of the
DGLAP and BFKL equations in the double logarithmic approximation,
\beq\label{d}
d(\tau)\,=\,\frac{(2\,\tau\,+\,1)}{\sqrt{4\,\tau\,+\,1}}\,
\ln\frac{\sqrt{4\,\tau\,+\,1}+1}{\sqrt{4\,\tau\,+\,1}-1}\,-\, 1.
\eeq
In derivation of \eq{dist} the phenomenological parametrization
of anomalous dimension in the Mellin moment variable $\omega$ was used
\cite{EKL}.
\beq\label{anom}
\gamma(\omega)\,=\,\alpha_s\,\left(\frac{1}{\omega}\,-\,1\right).
\eeq
The first term in the right-hand-side of \eq{anom} is the leading
order contribution to the anomalous dimension of the DGLAP and BFKL
equations in the double-logarithmic approximation. The second term is
the phenomenological correction which imposes momentum
conservation $\gamma(1)=0$ and describes DIS data well. We also
imposed the correct large $x$ behavior on  $\varphi_A$ by multiplying it
by a factor $(1-x)^4$. 

Let us consider the unintegrated gluon distribution in various
kinematic regions. First, consider a perturbative regime in which the collinear
factorization is valid $q_\bot \gg k_\mathrm{geom}(y^*)$, see \eq{coll} 
and \eq{qtc}.   
Performing expansion in the exponent of the Bessel function in this
regime, using its  well-known asymptotic behavior 
$I_0(z)\approx e^z$ up to a slowly varying logarithmic factors and
expanding  in \eq{d} $d(\tau)\approx Q_s^2(y^*)/q^2_\bot$ we get
\begin{eqnarray}\label{appr1}
\varphi_A(x,q^2_\bot) &\approx& \frac{4\,C_F}{2(2\pi)^2}
 \frac{S_A}{\as}\,\frac{Q_s^2(y_0^*)}
{q^2_\bot}
\left(\frac{Q_s^2(y^*)}{ q^2_\bot}\right)^{1.3\as}\nonumber\\
&& \times
\exp\left\{\left(\ln\frac{Q_s^2(y^*)}{Q_s^2(y_0^*)}\,\ln\frac{
 q^2_\bot}{Q_s^2(y_0^*)}\right)^{1/2}\right\}\,(1-x)^4.
\end{eqnarray}
Next, when the typical parton momentum in the heavy-ion wave function
approaches the saturation scale $Q_s(y)\lsim\, q_\bot\ll \kg(y)$
 one can expand the Bessel function as \cite{KLM}: 
\beq
\exp\left\{\left(\ln\frac{Q_s^2(y^*)}{Q_s^2(y_0^*)}\,\ln\frac{
q^2_\bot}{Q_s^2(y_0^*)}\right)^{1/2}\,-\,\ln\frac{q^2_\bot}{Q_s^2(y_0^*)}
\right\}\,
\approx\, (Q_s^2(y^*)/q^2_\bot)^{1/2},
\eeq 
which yields
\beq\label{appr2}
\varphi_A(x,\un q^2)\,\approx\, \frac{4\,C_F}{2(2\pi)^2}\frac{S_\bot}{\as}\,
\left(\frac{Q_s^2(y^*)}{q^2_\bot}\right)^{1.3\as\,+\, 1/2}\,(1-x)^4.
\eeq
Note that the anomalous dimension is $\gamma=1/2+\order{\as}$ as desired.
Finally, in the saturation region $q_\bot<Q_s(y)$ the unintegrated
gluon distribution reads
\beq\label{approx3}
\varphi_A(x,\un q^2)\,\approx\,\frac{4\,C_F}{2(2\pi)^2}\frac{S_\bot}{\as}\,
\ln (Q_s^2/q_\bot^2)\,(1-x)^4.
\eeq
This expression coincides with the formula for the 
unintegrated gluon distribution in the saturation regime in 
the quasi-classical approximation \cite{KKT}. Actually, it also holds 
beyond the quasi-classical approximation. Indeed, $\varphi_A(x,\un q^2)$ 
can be expressed via the gluon dipole forward scattering amplitude $N_G(x, 
\un r)$ as \cite{KTin,KKT}
\beq\label{FN}
\varphi_A(x,\un q^2)\,=\,\frac{4\,C_F\, S_A}{\as (2\pi)^2}
\int_0^\infty dr\, r\, J_0(kr)\frac{1}{r^2}\, N_G(x,\un r^2),
\eeq
where $r$ is the dipole transverse size. As was argued in \cite{LT}, the 
forward scattering amplitude in the momentum space at $x\rightarrow 0$ and 
$q$ fixed is 
\beq
N(x,\un q)\,=\,\ln(Q_s/q).
\eeq
In the configuration space we have 
\beq
N(x,\un r)\,=\, r^2\int_0^{Q_s} dk\, k\,
J_0(kr)\,\ln (Q_s/k)\,=\, 1\,-\, J_0(Q_sr).
\eeq
Using the unitarity constraint $N_G=2N-N^2$ and \eq{FN} we derive
\begin{eqnarray}
\varphi_A(x,\un q^2)&=& \frac{4\, C_F\, S_A}{\as\,  (2\pi)^2}\,
\int_0^\infty\frac{dr}{r} J_0(q r)\left( 1\,-\, J_0^2(Q_s r)\right)\\
&=&
\frac{4\, C_F\, S_A}{\as\, 2\, (2\pi)^2}\,
\ln(Q_s^2/\un q^2)\,+\mathcal{O}(q/Q_s),
\end{eqnarray}
which coincides with \eq{approx3}.

Dependence of the saturation scale on energy is known in the
double-logarithmic approximation \cite{LT,Iancu:2001md}. It was shown in
\cite{GBW} that the photon structure function in the DIS in the whole
kinematic region $x<0.01$ can be described by simple Glauber-like
parametrization of the gluon distribution with the following 
saturation scale
\beq\label{Qs0}
Q_s^2(x_{Bj})\,=\,\left(\frac{x_0}{x_{Bj}}\right)^\lambda,
\eeq
where $x_0$ and $\lambda$  are a certain empirical constants. 
In the case of heavy flavor production we substitute for $x_{Bj}$
expressions for $x$ and $y$ from \eq{xy1} and \eq{xy2} for each of the
nuclei:
\beq\label{QSC}
Q_s^{2\,c\bar c}(x)\,=\, \left(\frac{x_0}{x_1+x_2}\right)^\lambda
\eeq
and the same for the other nucleus (with $x\leftrightarrow y$).
Since $Q_s\sim A^{1/3}$ the saturation scales like
$N_\mathrm{part}^{1/3}$. In Ref.~\cite{KN} dependence of $Q_s$ on
$N_\mathrm{part}$ was calculated using the Glauber approach. Here we
are going to use the result of that calculation and refer the reader to 
the
\re{KN} for the explicit table. 

%%%%%%%%%%%%%%%%%%%%%%%%%%%%%%%%%%%%%%%%%%%%%%%%%%%%%%%%%%%%%%%%%%%%%%
\subsection{Numerical results}

%%%%%%%%%%%%
\begin{figure}
\begin{center}
\begin{tabular}{c}
\epsfig{file=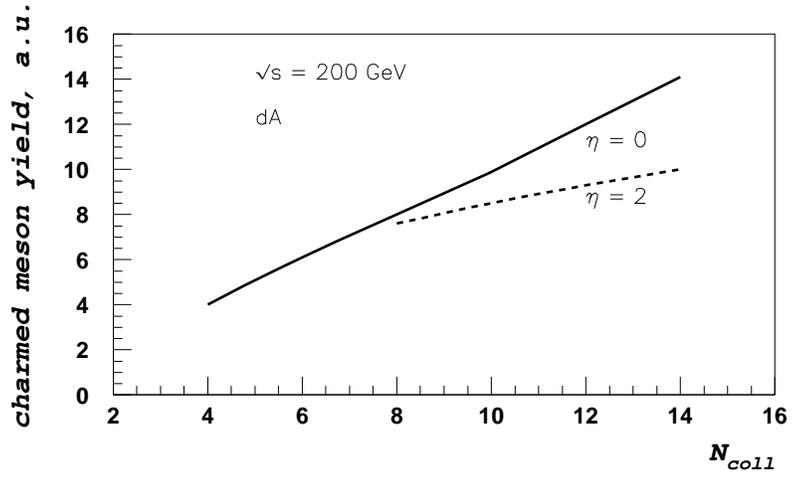, width=12cm}\\
\epsfig{file=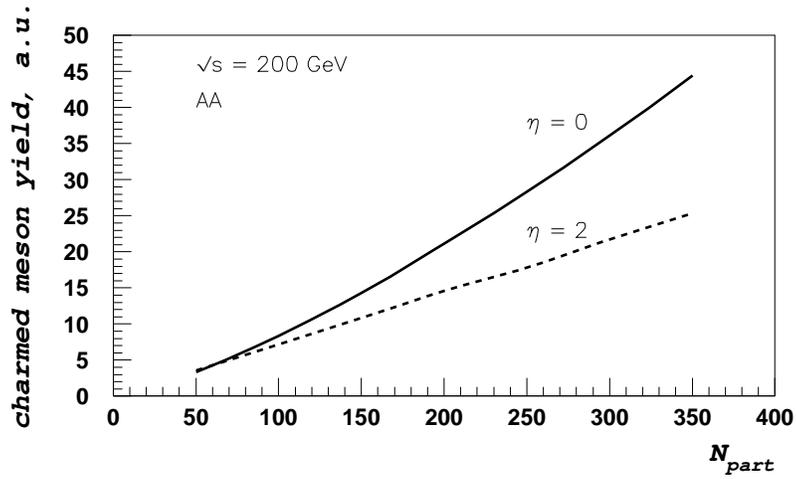,width=12cm}
\end{tabular}
\caption{Dependence of the charmed meson yield on centrality at midrapidity and 
pseudorapidity  $\eta=2$ for dAu and Au-Au collisions. }
\label{scaling1}
\end{center}
\end{figure}
%%%%%%%%%%%%
We perform numerical calculation of the charmed quark multiplicity 
spectrum in heavy-ion collisions at RHIC energies using  Eqs.~\eq{pp} 
and \eq{dist}. The parameters we use are:
charm quark mass $m=1.3$~GeV, the Golec-Biernat--W\"usthoff
parameter $\lambda=0.3$, the strong coupling $\as=0.35$ at $q_T\le 
Q_s$ and runs otherwise and $R_A=1.2\,\mathrm{fm}\,A^{1/3}$ for the 
nuclear radius. 
In \fig{scaling1} we present results of the calculation 
at midrapidity and in the forward region in $AA$ and $dA$ collisions at RHIC energy of $\sqrt{s}=200$ GeV 
as a function of centrality. We observe 
that at midrapidity quark spectra exhibit $N_\mathrm{coll}$ as 
expected since $Q_s(y^*=0)\sim m$ whereas at 
 forward rapidities $y^*> 2$ the  $N_\mathrm{part}$ and $\sqrt{N_\mathrm{part}^A}$ scaling is observed for $AA$ and $dA$ collisions, 
respectively. 

We conclude that the effects of multiple rescatterings and anomalous 
dimension variation do not affect the centrality dependence in the central rapidity region at RHIC energies for the charm quark 
production. However saturation 
significantly affects charm production mechanism in the forward rapidity 
region.

%%%%%%%%% fig %%%%%%%%%%%%
%\begin{figure}
%\begin{center}
%\epsfig{file=ccq.eps, width=13cm}
%\caption{The charmed quark spectrum in
%  heavy-ion collisions at  various energies. 
%}\label{ccq}
%\end{center}
%\end{figure}
%%%%%%%%%%%%%%%%%%%%%%%%
%The open charm spectrum in heavy -ion collisions is presented in 
%\fig{ccq}.
 The charmed-meson spectrum can be calculated
by plugging the unintegrated gluon distribution \eq{dist} 
into \eq{pp} and convolving with the charmed quark fragmentation
function $D_c^h(z)$ 
\beq\label{AAh}
\frac{d\sigma_{AA}^\mathrm{hadron}}{d^2  p_{1\bot}\, dy_1^*}\,=\,
\int_0^1 dz\, \frac{d\sigma_{AA}^\mathrm{jet}}{d^2 l_{1\bot}\, dy_1^*}\,
\theta \left(z\,-\,\frac{ p_\bot}{ q_\mathrm{max}}\right)\,
\delta\,( p_{1\bot}^2\,-\,z^2\, l_{1\bot}^2)\, D_c^h(z)\,\frac{1}{z^2}
\eeq
where $q_\mathrm{max}= \sqrt{S}/2$. In our calculation we use
the Peterson function \cite{Peterson} with $\epsilon=0.043$ \cite{PDG}. 
%%%%%%%%%%%%
\begin{figure}
\begin{center}
\epsfig{file=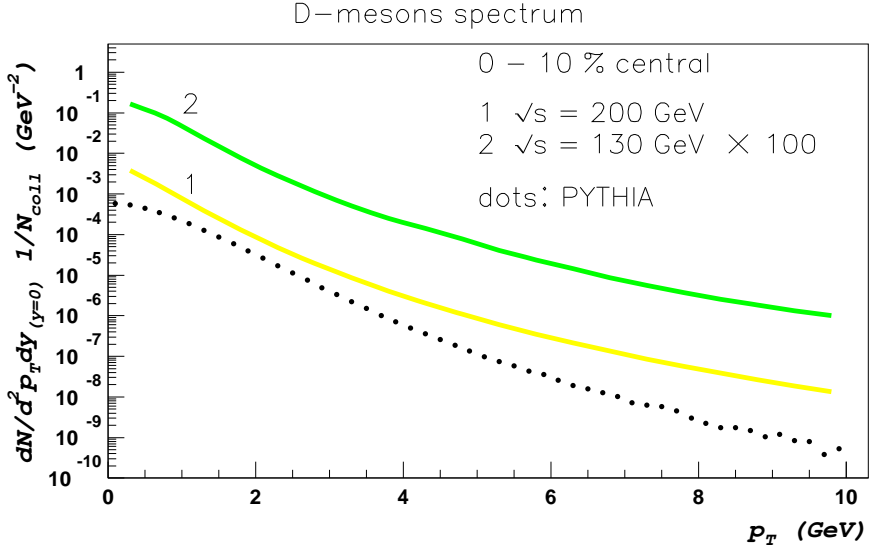, width=13cm}
\caption{The charmed meson spectrum in
  heavy-ion collisions at various energies. 
}\label{ccqd} 
\end{center}
\end{figure}
%%%%%%%%%%%
%
 Since we determined that the spectrum scales with 
$N_\mathrm{coll}$ in the midrapidity one can easily infer the $dA$ 
spectrum by scaling it with a corresponding number of binary collisions. 
The result is 
then compared to that obtained from PYTHIA event generator with  
the settings used by PHENIX collaboration \cite{PHENIX1}
(note that 
PYTHIA is based upon the collinear factorization).  In \fig{ccqd} we 
observe that PYTHIA spectrum is significantly softer than ours. 
Therefore, we predict much harder open charm spectrum. Of 
course, the steepness of the PYTHIA spectrum depends on the intrinsic momentum 
parameter $k_0$. Our spectrum can probably be reproduced by PYTHIA if 
one takes $k_0\simeq Q_s\gg \Lambda_\mathrm{QCD}$ and fixes the 
$K$-factor. This large value of intrinsic $k_T$ would however signal 
the breakdown of the collinear factorization approach.  

%%%%%%%%%%%%%%%%%%%%%%%
\subsection{Distribution in the total momentum of the heavy quark pair}

To get further insight into the dynamics of the heavy quark production it 
can be instructive to directly measure the total momentum of $q\bar q$ 
pair, which has to be equal to zero in the leading order pQCD calculation. 
Consider the relative motion of the two produced quark momenta
in the transverse plane.
In the region where collinear factorization is valid the transverse momentum
conservation of the process $gg\rightarrow c\bar c$ tells us that the
transverse momentum distribution must be proportional the 
 delta function $\delta(p_{1\bot} - p_{2\bot})$ in the center-of-mass
 frame. Due to hadronization the
delta function smears out to become a Gaussian of the typical width of
the order of $\Lambda^2$ independent of the energy and the
impact parameter (centrality) of the collision. At momenta $m_\bot\le
\kg\simeq Q_s$ the collinear factorization breaks down and we have
to use the $k_T$-factorization approach. The relevant process is
$gg\rightarrow ggc\bar c$. Therefore momentum conservation no longer
requires that the quark and antiquark momenta be correlated
back-to-back $p_{1\bot}=-p_{2\bot}$. In
general, the azimuthal distribution of quarks becomes asymmetric and
their transverse momentum distribution is disbalanced. The typical
difference between the quark momenta is $Q_s$ and therefore grows with
the energy of collision and centrality. It also become 
larger for quarks
produced far from midrapidity. As wee see, 
the direct consequence of the
saturation of the nuclei wave functions is that the total transverse
momentum of the quark--antiquark pair 
$(p_{1\bot}+p_{2\bot})^2$  does not vanish. 

Generally, in a given event a number $r$ of $q\bar q$
pairs is produced. Let $P_\bot$ be the total transverse momentum of
all quarks and anti quarks produced in a given event. The averaged
total transverse momentum in events with a given $r$ is
given by 
\begin{eqnarray}\label{Pdef}
\overline {(P_\bot^2)_r} &=&
\int\frac{dN^\mathrm{c\bar c}}
{d^2 p_{\bot 1}\, dy_1^* \ldots d^2p_{\bot 2r}\, dy_{2r}^*}
\left(\sum_{i=1}^{2r}p_{i\bot}\right)^2\prod_{i=1}^{2r} 
d^2 p_{\bot i}\, dy_i^*\nonumber\\
&&\times\left(\int\frac{dN^\mathrm{c\bar c}}{d^2 p_{\bot 1}\, dy_1^*\ldots
d^2 p_{\bot 2r}\, dy_{2r}^*}\right)^{-1}
\end{eqnarray}
With a good accuracy  quarks and antiquarks 
are produced in
pairs, thus at this approximation we can write
\begin{eqnarray}
\overline {(P_\bot^2)_r} &=& r\,r^{r-1}\,C^2_{2r}
\int\frac{dN^\mathrm{c\bar c}}
{d^2 p_{\bot 1}\, dy_1^* d^2p_{\bot 2}\, dy_2^*}
(p_{1\bot}+p_{2\bot})^2\, d^2 p_{\bot 1} dy_1^*\, d^2 p_{\bot 2}\,
dy_2^*\, \frac{1}{r\,C^2_{2r}}\nonumber\\
&=& r\, \overline {(P_\bot^2)_1}.
\end{eqnarray}
It is convenient to define the event-averaged total transverse
momentum of the $c\bar c$ pair as follows
\beq\label{eva}
\langle P_\bot^2\rangle\,=\,
\left\langle\, \frac{1}{r}\,\overline {(P_\bot^2)_1}
\right\rangle_\mathrm{all\; events}.
\eeq
We can estimate $\langle P_\bot^2\rangle$ using equations
\eq{obl1},\eq{obl2},\eq{obl3} together with \eq{thum1},\eq{thum2} and
\eq{thum3}. Since in the midrapidity region of RHIC $c\bar c$ is
produced perturbatively we find
\beq    
\langle P_\bot^2\rangle\,\simeq\,0, \quad y^*\simeq 0,\quad \mathrm{RHIC}.
\eeq
At larger collision energies, when  $Q_s(y^*=0)\gg m$ both nuclei are
saturated and we get (modulo logarithms)
\beq
\langle P_\bot^2\rangle\,\simeq\,Q^2_s, 
\quad y^*\simeq 0,\quad \mathrm{LHC}.
\eeq
At large rapidities, when only one of the nuclei is in saturation we
find
\beq
\langle P_\bot^2\rangle\,\simeq\,Q^2_{s\mathrm{\max}}. 
\quad |y^*|\gg 1,\quad \mathrm{RHIC,\;LHC}.
\eeq

%%%%%%%%%%%%%%%%%%%%%%%%%%%%%%%%%%%%%%%%%%%%%%%%%%%%%%%%%%%%%%%%%%%%%%
\section{Quenching of the heavy quark spectra \\ in QCD matter}

Our discussion so far has neglected the final state effects. 
However if hot QCD  matter is formed in heavy-ion collisions it could 
strongly influence the open charm spectrum, similarly to the case of light 
partons \cite{GW,BDMPS,WS}. The energy loss of heavy quarks however is different in 
one important aspect \cite{DK}: since the velocity of heavy quark is smaller 
than unity, the angular distribution of the gluons emitted in the medium 
vanishes in the forward direction (``dead cone'' effect). Consequently, 
the collinear singularities disappear, and the sensitivity of the result to
the infrared cutoff is greatly diminished which allows for a 
rigorous perturbative QCD treatment even at small transverse momenta. The resulting energy 
loss is significantly smaller than for light partons. 
Recently this issue was also examined in Refs \cite{HQloss}.

\begin{figure}
\begin{center}
\epsfig{file=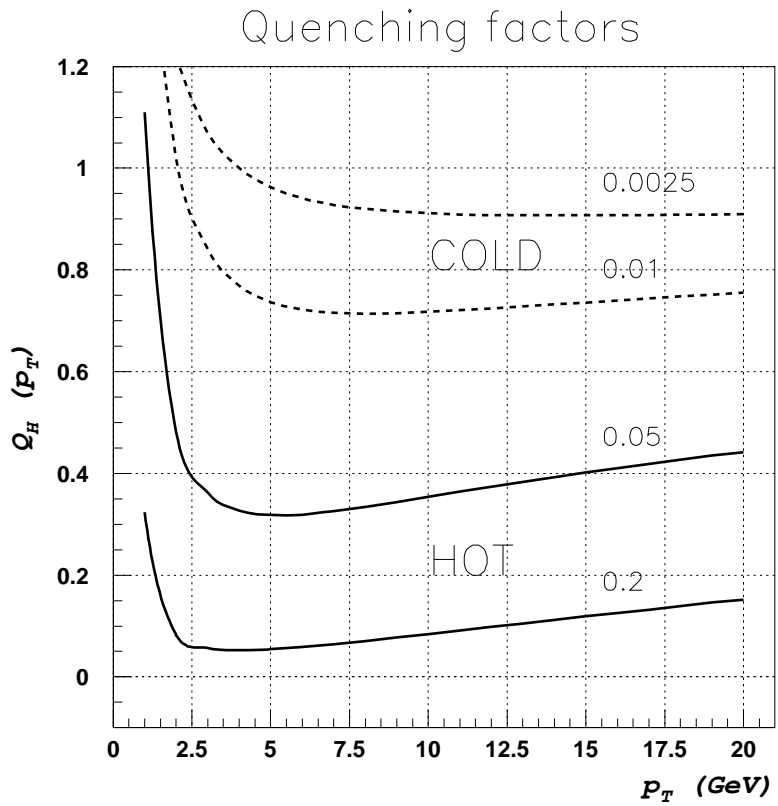, width=12cm}
\end{center}
\caption{Quenching factors for hot and cold media 
$\hat q = 0.2,\, 0.05,\, 0.01,\, 0.0025$~GeV$^3$. $L=5$~fm. 
}\label{quench}
\end{figure}

The medium-modified spectrum can be written as 
\cite{BDMPS,DK}
\beq\label{med}
\frac{d\sigma^\mathrm{med}}{dp_\bot^2}\,=\,
\frac{d\sigma^\mathrm{vac}}{dp_\bot^2}(p_\bot)\,Q_H(p_\bot),
\eeq
provided that the energy loss is much smaller than $p_\bot$. The quantity 
$d\sigma^\mathrm{med}/dp_\bot^2$ corresponds to the quark spectrum in the vacuum.
The quenching factor $Q_H(p_\bot)$ for heavy quarks has been evaluated as \cite{DK}
\beq\label{Q}
Q_H(p_\bot)\simeq \exp\left[
-\frac{2\as\, C_F}{\sqrt{\pi}}\,L\,
\sqrt{\hat q\,\frac{\mathcal{L}(p_\bot)}{p_\bot} }\,+\,
\frac{16\as}{9\sqrt{3}}\, L\, \left( \frac{\hat q\, m^2}{m^2+p_\bot^2} 
\right)^{1/3}\right],
\eeq
where 
\beq\label{incr}
\mathcal{L}(p_\bot)\,=\,-\, \frac{d}{d\ln p_\bot}\ln\left[ 
\frac{d\sigma^\mathrm{vac}}{dp_T}(p_T)
\right],
\eeq
$\hat q$ is the transport coefficient and $L$ is the medium size.
The charmed meson spectrum in vacuum calculated in the previous 
section has intrinsic momentum $p_0\simeq Q_s$.
%\beq\label{param}
%\frac{d\sigma}{dp_\bot^2\, 
%dy^*}\,=\,\frac{C\,S_A}{\left(p_0^2\,+\,p_\bot^2\right)^{n/2}},
%\eeq
Obviously at $p_\bot\gg p_0$ we have 
$\mathcal{L}=\mathrm{const}$ and the quenching factor rapidly approaches 
unity. However, when $p_\bot\ll  p_0$ we have $\mathcal{L}\propto p_\bot$ 
and the first term in the exponent in \eq{Q} stays almost constant. This   
implies much slower dependence of the quenching factor  $Q_H(p_\bot)$ 
on $p_T$. We calculate  $\mathcal{L}$ by 
substituting the quark spectrum into \eq{incr}. Then, using \eq{Q} we 
calculate the quenching factor for AA 
collisions (hot matter) and dA collisions (cold medium) which is plotted in 
\fig{quench}.   
%%%%%%%%%%%%
%%%%%%%%%%%%
Note, that the spectrum shape in vacuum in $k_T$-factorization is much 
harder than that in the parton model. Therefore, the fact that we use 
$k_T$-factorization is both qualitatively and quantitatively essential for the result 
presented in 
\fig{quench}. Even though \fig{quench} exhibits substantial quenching for the 
values of the gluon transport coefficient that we used, it is still significantly smaller 
than the quenching factor for the light quarks computed under the same assumptions about 
the density of the medium.

%%%%%%%%%%%%%%%%%%%%%%%%%%%%%%%%%%%%%%%%%%%%%%%%%%%%%%%%%%%%%%%%%%%%%%
\section{Summary}

In conclusion, we have calculated the charmed meson spectrum for heavy
ion collisions in the framework of the Color Glass Condensate. 
 We found 
that at midrapidity the spectrum at RHIC energies scales as 
$N_\mathrm{coll}$, while in the forward region it shows 
$N_\mathrm{part} (\sqrt{N_\mathrm{part}^A})$ scaling in AA (dA) collisions. Our results are thus different from 
the predictions based on collinear factorization, even after the leading twist 
shadowing is included. 

Using the explicit form of the charm quark spectrum computed in this approach,  we have calculated the 
quenching factors due to interaction of quarks with hot and cold media. 
Although at very high $p_T$ the quenching factor approaches unity, in a 
wide region of $2< p_T  < 15$~GeV it stays almost flat. This feature of the 
quenching factor is a direct consequence of the saturation of the nuclear 
wave functions which  brings in the dimensionful scale $p_0\sim Q_s$.
 
We have suggested to determine the saturation scale by 
measuring the total transverse 
momentum of charmed mesons produced in a given
rapidity interval. We expect that unlike in naive leading order perturbation theory, 
the total transverse momentum of the pair will not vanish. On the contrary, it will grow with the energy,
rapidity and centrality in the same way as the saturation scale.

%%%%%%%%%%%%%%%%%%%%%%%%%%%%%%%%%%%%%%%%%%%%%%%%%%%%%%%%%%%%%%%%%%%%%%%%%%%%

\vskip0.3cm
{\large\bf Acknowledgments}
%\vskip0.3cm

The authors are indebted to Yuri Dokshitzer, Yuri Kovchegov, Eugene Levin, Larry McLerran, 
Raju Venugopalan and  Xin-Nian Wang for fruitful discussions of various aspects of this work.
We thank Stefan Kretzer for helpful discussion of heavy quark 
fragmentation functions, and Zhangbu Xu for providing us with the PYTHIA results for charmed meson spectra. 
The research of D.K. and K.T. was supported by 
the U.S. Department of Energy under Grant No. DE-AC02-98CH-10886. 
The work of K. T. was also sponsored in part by the U.S. Department of 
Energy under Grant No. DE-FG03-00ER41132.

%%%%%%%%%%%%%%%%%%%%%%%%%%%%%%%%%%%%%%%%%%%%%%%%%%%%%%%%%%%%%%%%%%%%%%%%%%%
\section*{Appendix}\appendix

In the Appendix we cite the heavy quark production 
amplitude $\mathcal{A}(gg\rightarrow ggc\bar c)$ as given in
\cite{CCH}.

\beq\label{amp}
\mathcal{A}(gg\rightarrow ggc\bar c)\,=\,\frac{1}{2N_c}\,
\mathcal{A}^\mathrm{ab}\,+\,
\frac{N_c}{2(N_c^2-1)}\,\mathcal{A}^\mathrm{nab},
\eeq
where 
\beq\label{ab}
\mathcal{A}^\mathrm{ab}\,=\,
x^2y^2S^2\left\{
\frac{1}{(t-m^2)(u-m^2)}\,-\,\frac{1}{q_{1\bot}^2q_{2\bot}^2}
\left[1\,+\,\frac{x_2y_1S}{t-m^2}\,+\,
\frac{x_1y_2S}{u-m^2}\right]^2\right\},
\eeq
\begin{eqnarray}
\mathcal{A}^\mathrm{nab}&=&
x^2y^2\left\{ S^2\left[-\frac{1}{(t-m^2)(u-m^2)}\,-\,
\frac{1}{s}\left(\frac{1}{t-m^2}\,-\,\frac{1}{u-m^2}\right)
\left(\frac{x_1}{x}\,-\,\frac{y_1}{y}\right)\right.\right.
\nonumber\\
&&+
\left.\left.\frac{2}{ S\,s\, xy}\right]\,+\,
\left[\frac{S}{2}\,+\,S^2\frac{x_2y_1}{t-m^2}\,-\,\frac{\Delta}{s}\right]
\left[\frac{S}{2}\,+\,S^2\frac{x_1y_2}{u-m^2}\,+\,\frac{\Delta}{s}\right]
\right\}.\label{nab}
\end{eqnarray}
\beq\label{Delta}
\Delta\,=\, S\left[
  Sx_1y_2\,-\,Sx_2y_1\,+\,q_{1\bot}^2\frac{x_2}{x}\,-\,
q_{2\bot}^2\frac{y_2}{y}\,+\,\frac{1}{2}(u-t)\,+\,
\frac{1}{2}(q_{2\bot}^2\,-\,q_{1\bot}^2)\right]
\eeq
\beq
t\,=\, (q_1-p_2)^2,\quad u\,=\,(q_1-p_1)^2,\quad
s\,\equiv\,\mathcal{M}^2\,=\, (p_1+p_2)^2,
\eeq 
\beq
q_1\,=\,y\,p_A\,+\,q_{1\bot},\quad q_2\,=\,x\,p_B\,+\,q_{2\bot}
\eeq
\beq
p_{1,2}\,=\,y_{1,2}\,p_A\,+\,x_{1,2}\,p_B\,+\,p_{1,2\bot},
\eeq
where $2 p_A\cdot p_B=S$.

%%%%%%%%%%%%%%%%%%%%%%%%%%%%%%%%%%%%%%%%%%%%%%%%%%%%%%%%%%%%%%%%%%%%%%%%%%%%

\end{document}